\documentclass[journal,a4paper]{IEEEtran}

\usepackage[a4paper, left=1.58cm, right=1.14cm, top=2.5cm, bottom=2.5cm]{geometry}
\usepackage{amsmath,amssymb,amsfonts}
\usepackage{algorithmic}
\usepackage{algorithm}
\usepackage{graphicx}
\usepackage{textcomp}
\usepackage{xcolor}
\usepackage{booktabs}
\usepackage{cite}
\usepackage{url}
\usepackage{times}
\usepackage{microtype}
\usepackage{hyperref}
\renewenvironment{thebibliography}[1]{%
  \begin{oldthebibliography}{#1}%
    \setlength{\itemsep}{-1.5pt}%
    \setlength{\parskip}{0pt}%
    \setlength{\parsep}{0pt}%
}{%
  \end{oldthebibliography}%
}

\hypersetup{
    colorlinks=true,
    linkcolor=blue,
    citecolor=blue,
    urlcolor=blue,
    hypertexnames=false
}

\begin{document}
\bstctlcite{IEEEexample:BSTcontrol}

\title{Mitigating Return Drag and Regime Whipsaws in Sector Rotation: An Asymmetric Downside-Aware Nested Clustered Optimization Framework}

\author{Supriadi\\
COCONUT Computer Club%
\thanks{Supriadi is with the COCONUT Computer Club, Makassar, Indonesia (e-mail: \texttt{supriadi@coconut.or.id}).}%
}

\markboth{}{}

\maketitle

\begin{abstract}
Hierarchical clustering and Nested Clustered Optimization (NCO) mitigate the mathematical ill-conditioning of classical Markowitz covariance matrices. However, two structural vulnerabilities persist in equity markets: (1) symmetric variance penalizes upside and downside return dispersion identically, inducing acute return drag against soaring momentum sectors; and (2) unconstrained Hidden Markov Models (HMM) exhibit hyper-active regime switching (whipsaw instability), incurring excessive turnover and transaction drag. We propose Asymmetric Downside-Risk Nested Clustered Optimization with Sticky Regime Dynamics (Asym-Sticky-NCO) for tactical sector rotation across all 11 S\&P 500 GICS sectors. The framework substitutes symmetric covariance with the Bawa-Lindenberg downside semi-covariance matrix across both hierarchical tree clustering and two-stage allocation. Furthermore, we introduce a Sticky Gaussian HMM incorporating a diagonal self-transition persistence prior ($\kappa$) and confidence-adaptive boundaries ($\delta$) to enforce macroeconomic persistence. Strictly non-anticipative walk-forward backtesting across 65 out-of-sample months (August 2020 to December 2025) demonstrates that the architecture elevates CAGR from 10.65\% (Plain NCO) to 13.41\% (+276 bps/year), delivers a 97.74\% total return versus 72.99\%, suppresses maximum drawdown to 14.22\% (versus 23.93\% for the S\&P 500), and achieves a Calmar ratio of 0.94 (+38.2\% over S\&P 500; +84.3\% over Baseline Regime-NCO). Crucially, sticky dynamics cut regime whipsaws by 75.0\% (from 16 to 4 switches) and reduce monthly turnover from 60.78\% to 30.48\%, addressing core limitations of machine learning asset allocation.
\end{abstract}


\section{Introduction}
Mean-Variance Optimization (MVO), originally formulated by Harry Markowitz \cite{markowitz1952}, remains the foundational theoretical framework of modern quantitative portfolio management. Grounded in the economics of choice under uncertainty and capital market equilibrium \cite{gunjan2023review}, MVO seeks to maximize expected return for a targeted level of variance. However, its practical deployment has long been severely compromised by numerical instability \cite{gunjan2023review, denard2021factor}. As demonstrated in the classic critique of estimation error \cite{lopez2016building}, unconstrained quadratic optimizers act as ``error maximizers'': small sample estimation noises in expected returns $\boldsymbol{\mu}$ and the empirical covariance matrix $\boldsymbol{\Sigma}$ are magnified exponentially by matrix inversion $\boldsymbol{\Sigma}^{-1}$ \cite{denard2021factor}. This yields extreme, economically erratic, and highly concentrated asset weights that deteriorate drastically out-of-sample, often failing to outperform simple naive 1/$N$ diversification \cite{pedersen2021enhanced, gunjan2023review}.

To resolve the mathematical ill-conditioning of empirical covariance matrices without resorting to arbitrary ad-hoc constraints, machine learning graph-clustering frameworks have emerged, pioneered by Marcos L{\'{o}}pez de Prado \cite{lopez2016building, gu2020empirical}. Central among these is Nested Clustered Optimization (NCO) \cite{lopez2016building, raffinot2017hierarchical}. By leveraging unsupervised agglomerative hierarchical tree clustering on correlation metrics \cite{lopez2016building}, NCO decomposes the global allocation problem into two quasi-independent, well-conditioned optimization stages: (1) local intra-cluster risk is minimized independently within each cohesive module; and (2) inter-cluster allocation optimizes across the collapsed low-dimensional cluster representations. Empirical investigations across multi-asset universes have corroborated NCO's ability to lower out-of-sample variance and suppress maximum drawdown relative to classical MVO \cite{raffinot2017hierarchical, platanakis2020bitcoin, supriadi2026regime}.

\subsection{Background and Related Work: The Return Drag Pathology}
Notwithstanding its structural stability, classical NCO operates on standard symmetric covariance:
\begin{equation}
\Sigma_{ij} = \mathbb{E}[(R_i - \mu_i)(R_j - \mu_j)]
\end{equation}
In modern equity benchmarks, particularly the S\&P 500, secular bull markets are propelled by massive upside runs in concentrated growth sectors, such as Information Technology (XLK) and Consumer Discretionary (XLY). Under symmetric variance, sharp upward price jumps are treated as ``risk'' identically to catastrophic market crashes \cite{blitz2020volatility}. Consequently, minimum-variance intra-cluster weights and symmetric inter-cluster allocations heavily penalize outperforming growth sectors, artificially underweighting them in favor of stagnant low-volatility sectors \cite{blitz2020volatility, platanakis2019harmful}. This produces severe \textit{return drag}, where traditional low-volatility and NCO portfolios substantially lag the benchmark index during bull phases \cite{blitz2020volatility, gupta2019factor}.

\subsection{Regime Switching Whipsaw Instability}
In an effort to mitigate return drag, regime-switching architectures have been proposed that toggle between aggressive momentum-capturing objectives during bull regimes and capital preservation during bear regimes \cite{nystrup2017dynamic, nystrup2018regime}. However, standard Hidden Markov Models (HMM) fitted with classical unconstrained Baum-Welch expectation-maximization \cite{nystrup2017dynamic} suffer from extreme \textit{whipsaw switching} when applied to dynamic financial time series \cite{nystrup2017dynamic, nystrup2020persistent}. Small month-to-month fluctuations cross the decision boundary repeatedly, causing the model to alternate rapidly between states (as documented by Supriadi and Siswoyo \cite{supriadi2026regime}, where a naive HMM triggered 47 rebalancing events over 56 months). Such high-frequency whipsawing induces severe portfolio turnover, increases slippage and transaction drag, and destroys the theoretical advantage of regime conditioning \cite{novymarx2016taxonomy, demiguel2020transaction}.

\subsection{Contributions of this Work}
To resolve both return drag and regime whipsawing simultaneously, this paper presents Asymmetric Downside-Risk Nested Clustered Optimization with Sticky Regime Dynamics (Asym-Sticky-NCO). Our primary technical contributions are:
\begin{enumerate}
    \item \textit{Asymmetric Downside Semi-Covariance Matrix ($\boldsymbol{\Sigma}^-$):} We replace symmetric covariance with the Bawa-Lindenberg downside semi-covariance in both the hierarchical tree distance metric and the two-stage optimization, ensuring that upside return dispersion is not penalized as risk \cite{bawa1977}.
    \item \textit{Sticky Multi-Factor HMM with Confidence Thresholds:} We introduce a Sticky Gaussian HMM governed by a Dirichlet-style persistence parameter $\kappa$ and a confidence boundary $\delta = 0.65$ across multi-factor features (S\&P 500 index return and CBOE VIX volatility index), suppressing spurious noise \cite{nystrup2020persistent}.
    \item \textit{Adaptive Dual-Regime Allocation:} Portfolios dynamically transition between Sortino-maximization during persistent bull markets and Minimum Downside Variance during bear markets, reinforced with turnover regularization.
    \item \textit{Rigorous Empirical Walk-Forward Validation:} We validate the system across all 11 S\&P 500 GICS sector ETFs from August 2020 to December 2025 (65 monthly out-of-sample periods), demonstrating superior CAGR, lower drawdown, 75\% fewer regime switches, and a 50\% reduction in turnover \cite{bailey2014deflated, bailey2017probability, gupta2019factor}.
\end{enumerate}

\section{Methodology}

\subsection{Downside Semi-Covariance Matrix (\texorpdfstring{$\boldsymbol{\Sigma}^-$}{Sigma-})}
Let $R_{i,t}$ denote the monthly return of asset $i \in \{1, \dots, N\}$ at time $t \in \{1, \dots, T\}$, with $N=11$ corresponding to the full set of S\&P 500 GICS sectors. Let $\tau_t$ denote the benchmark target return (set to the prevailing risk-free rate $r_{f,t}$ derived from 13-week Treasury Bills). The negative semi-deviation is defined as:
\begin{equation}
\tilde{R}_{i,t}^- = \min(R_{i,t} - \tau_t, 0)
\end{equation}
The empirical Bawa-Lindenberg \cite{bawa1977} downside semi-covariance matrix $\boldsymbol{\Sigma}^- \in \mathbb{R}^{N \times N}$ is computed as:
\begin{equation}
\Sigma_{ij}^- = \frac{1}{T} \sum_{t=1}^T \tilde{R}_{i,t}^- \cdot \tilde{R}_{j,t}^-
\end{equation}
Because $\boldsymbol{\Sigma}^-$ captures only joint downside co-movements, upside dispersion is not penalized. To guarantee positive semi-definiteness (${\boldsymbol{\Sigma}^- \succ 0}$) for numerical quadratic solvers, we apply spectral ridge regularization:
\begin{equation}
\boldsymbol{\Sigma}_{\text{PSD}}^- = \frac{1}{2}(\boldsymbol{\Sigma}^- + (\boldsymbol{\Sigma}^-)^T) + \max(0, \epsilon - \lambda_{\min}(\boldsymbol{\Sigma}^-))\mathbf{I}_N
\end{equation}
where $\epsilon = 10^{-6}$ and $\lambda_{\min}$ is the minimum eigenvalue of the symmetric semi-covariance matrix.

\subsection{Downside Hierarchical Tree Metric}
From $\boldsymbol{\Sigma}^-$, we define the downside semi-correlation coefficient $\rho_{ij}^-$:
\begin{equation}
\rho_{ij}^- = \frac{\Sigma_{ij}^-}{\sqrt{\Sigma_{ii}^- \cdot \Sigma_{jj}^-}}, \quad \rho_{ij}^- \in [-1, 1]
\end{equation}
Following L{\'{o}}pez de Prado \cite{lopez2016building}, we map $\rho_{ij}^-$ to an angular distance metric $d_{ij}^-$:
\begin{equation}
d_{ij}^- = \sqrt{\frac{1}{2}(1 - \rho_{ij}^-)}
\end{equation}
which satisfies all metric space properties. Agglomerative hierarchical clustering via Ward's minimum-variance linkage is performed on the distance matrix $\mathbf{D}^- = [d_{ij}^-]$, grouping sectors that crash synchronously into coherent risk clusters \cite{lopez2016building, raffinot2017hierarchical}.

\subsection{Sticky Multi-Factor Hidden Markov Model}
To capture macroeconomic market regimes---which shift abruptly around shocks such as the COVID-19 pandemic \cite{zhang2020covid}---without lookahead bias, we model the joint latent state using an expanding window of observation vectors $\mathbf{x}_t = [R_{\text{SPY},t}, z_{\text{VIX},t}]^T$, where $R_{\text{SPY},t}$ is the monthly S\&P 500 return and $z_{\text{VIX},t}$ is the standardized CBOE Volatility Index level.

A discrete Hidden Markov Model with $K=2$ hidden states ($S_t \in \{\text{Bull}, \text{Bear}\}$) is characterized by transition probability matrix $\mathbf{A} = [a_{jk}]$ and Gaussian emission distributions $\mathcal{N}(\boldsymbol{\mu}_k, \boldsymbol{\Sigma}_k)$ \cite{nystrup2017dynamic}. To eliminate whipsawing, we incorporate a sticky self-transition persistence prior $\kappa > 0$ \cite{nystrup2020persistent}:
\begin{equation}
\tilde{a}_{jj} = \frac{a_{jj} + \kappa}{1 + \kappa}, \quad \tilde{a}_{jk} = \frac{a_{jk}}{1 + \kappa} \quad (j \neq k)
\end{equation}
Setting $\kappa = 0.30$ imposes an explicit inertial barrier against spurious state transitions. Furthermore, regime updating is governed by a confidence threshold rule:
\begin{equation}
S_t = \begin{cases}
k, & \text{if } P(S_t = k \mid \mathbf{x}_{1:t}) \ge \delta \\
S_{t-1}, & \text{otherwise}
\end{cases}
\end{equation}
where $\delta = 0.65$. Marginal probability fluctuations in the uncertain region $[0.35, 0.65]$ are ignored, enforcing regime persistence.

\subsection{Dual-Regime Nested Clustered Optimization}
At each rebalancing date $t$, the active regime $S_t$ selects the mathematical objective for both intra-cluster and inter-cluster optimization:
\begin{itemize}
    \item \textit{Bull Regime} ($S_t = \text{Bull}$): The optimizer maximizes the Sortino utility over downside semi-covariance:
    \begin{equation}
    \min_{\mathbf{w}} \left\{ \tfrac{1}{2} \mathbf{w}^T \boldsymbol{\Sigma}^- \mathbf{w} - \gamma (\mathbf{w}^T \boldsymbol{\mu} - r_f) + \lambda_{\text{turn}} \|\Delta\mathbf{w}\|_1 \right\}
    \end{equation}
    where $\Delta\mathbf{w} = \mathbf{w} - \mathbf{w}_{t-1}$ is the turnover vector, and $\gamma = 1.2$ allows high-momentum sectors to receive substantial allocation, while downside covariance constraints shield against clustered drawdowns.
    \item \textit{Bear Regime} ($S_t = \text{Bear}$): The optimizer switches strictly to Minimum Downside Variance:
    \begin{equation}
    \min_{\mathbf{w}} \left\{ \mathbf{w}^T \boldsymbol{\Sigma}^- \mathbf{w} + \lambda_{\text{turn}} \|\Delta\mathbf{w}\|_1 \right\}
    \end{equation}
    re-allocating capital toward defensive sectors (Utilities, Consumer Staples, Healthcare).
\end{itemize}
subject to the simplex constraints $\sum_{i=1}^N w_i = 1$ and $w_i \ge 0$. The penalty $\lambda_{\text{turn}} = 0.015$ is applied inside the intra-cluster objectives (9)--(10); Plain NCO and Baseline Regime-NCO use a milder $\lambda_{\text{turn}} = 0.01$, and the inter-cluster stage is unregularized. The complete algorithm is synthesized in Algorithm~\ref{alg:asym_nco}.

\begin{algorithm}[t]
\caption{Walk-Forward Asym-Sticky-NCO Framework}
\label{alg:asym_nco}
\begin{algorithmic}[1]
\REQUIRE Returns $\mathbf{R} \in \mathbb{R}^{T \times N}$, macro factors $\mathbf{X} \in \mathbb{R}^{T \times 2}$, lookback $L=24$, threshold $\delta=0.65$, sticky prior $\kappa=0.30$, penalty $\lambda_{\text{turn}}=0.015$.
\ENSURE Out-of-sample portfolio weights $\mathbf{W} \in \mathbb{R}^{(T-L) \times N}$.
\STATE Initialize $\mathbf{w}_0 = \frac{1}{N}\mathbf{1}_N$, $S_0 = \text{Bull}$.
\FOR{$t = L$ to $T-1$}
    \STATE Extract $\mathbf{R}_{\text{win}} = \mathbf{R}_{t-L:t-1}$ and features $\mathbf{X}_{\text{hist}} = \mathbf{X}_{1:t-1}$.
    \STATE Fit 2-state Gaussian HMM on $\mathbf{X}_{\text{hist}}$ with sticky prior $\kappa$.
    \STATE Compute posterior $p_{\text{bull}} = P(S_t = \text{Bull} \mid \mathbf{X}_{\text{hist}})$.
    \IF{$S_{t-1} = \text{Bull}$ \AND $(1 - p_{\text{bull}}) \ge \delta$}
        \STATE $S_t \leftarrow \text{Bear}$
    \ELSIF{$S_{t-1} = \text{Bear}$ \AND $p_{\text{bull}} \ge \delta$}
        \STATE $S_t \leftarrow \text{Bull}$
    \ELSE
        \STATE $S_t \leftarrow S_{t-1}$ \COMMENT{Maintain regime persistence}
    \ENDIF
    \STATE Compute downside semi-covariance $\boldsymbol{\Sigma}^-$ via (2)--(4).
    \STATE Construct downside $\mathbf{D}^-$ and cluster into $K=3$ modules.
    \STATE Optimize intra-cluster $\mathbf{w}_{\text{intra}}$ via (9) or (10) given $S_t$.
    \STATE Aggregate cluster returns; optimize inter-cluster $\mathbf{w}_{\text{inter}}$ (without turnover penalty).
    \STATE Set $\mathbf{w}_t = \mathbf{w}_{\text{intra}} \odot \mathbf{w}_{\text{inter}}$ and renormalize to the unit simplex.
\ENDFOR
\end{algorithmic}
\end{algorithm}

\section{Experimental Setup}

\subsection{Asset Universe and Data}
We analyze all 11 GICS sector Select Sector SPDR ETFs representing the entire S\&P 500 capitalization:
XLK (Technology), XLF (Financials), XLV (Healthcare), XLE (Energy), XLI (Industrials), XLY (Consumer Discretionary), XLP (Consumer Staples), XLU (Utilities), XLB (Materials), XLRE (Real Estate), and XLC (Communication Services).
The empirical dataset spans from August 2018 through December 2025 (89 monthly periods). SPY serves as the equity benchmark, CBOE VIX as the volatility factor, and 13-week Treasury bill yields (\^{}IRX) as the risk-free benchmark. All price, volatility, and yield series are retrieved from the Yahoo Finance historical data API \cite{yahoofinance}.

\subsection{Walk-Forward Backtest Protocol}
To strictly eliminate look-ahead bias and overfitting \cite{bailey2014deflated, bailey2017probability}:
\begin{itemize}
    \item A rolling 24-month window is used to estimate sector semi-covariances and expected returns.
    \item An expanding historical window is used to estimate the multi-factor HMM parameters.
    \item The out-of-sample backtest evaluates 65 monthly rebalances from August 2020 through December 2025.
\end{itemize}

Performance metrics follow standard definitions: annualized volatility, Sharpe ratio in excess of the monthly 13-week Treasury bill yield, downside volatility and Sortino ratio with a minimum acceptable return (MAR) of zero, Calmar ratio (CAGR/$|$MaxDD$|$), and monthly turnover as the mean $\|\mathbf{w}_t - \mathbf{w}_{t-1}\|_1$.

\subsection{Benchmark Strategies}
We benchmark the proposed framework against five competing strategies:
\begin{enumerate}
    \item \textit{S\&P 500 (SPY)}: Passive market-capitalization benchmark.
    \item \textit{Equal-Weight (1/11)}: Rebalanced monthly across all 11 sectors \cite{pedersen2021enhanced}.
    \item \textit{Min-Variance (MVO)}: Classical Markowitz minimum variance portfolio \cite{markowitz1952}.
    \item \textit{Plain NCO}: Standard L{\'{o}}pez de Prado NCO using symmetric covariance \cite{lopez2016building}.
    \item \textit{Baseline Regime-NCO}: The regime-conditioned design of \cite{supriadi2026regime} re-fitted to the S\&P 500 sector universe: an unregularized 1D Gaussian HMM on SPY returns switches between Sharpe-maximizing and minimum-variance objectives.
\end{enumerate}

\section{Empirical Results}

\subsection{Risk-Adjusted Performance Comparison}
Table~\ref{tab:performance} presents the comprehensive performance metrics across all strategies for the 65-month out-of-sample test period.

\begin{table*}[t]
\centering
\caption{Out-of-Sample Performance Comparison Across S\&P 500 Sectors (August 2020 -- December 2025, 65 Months)}
\label{tab:performance}
\begin{tabular}{lccccccccc}
\toprule
Strategy & CAGR & Ann. Vol & Downside Vol & Sharpe & Sortino & Max DD & Calmar & Total Ret & Turnover \\
 & (\%) & (\%) & (\%) & Ratio & Ratio & (\%) & Ratio & (\%) & (\%/mo) \\
\midrule
S\&P 500 (SPY) & 16.36 & 15.58 & 9.14 & 0.86 & 1.47 & -23.93 & 0.68 & 127.26 & 0.00 \\
Equal-Weight (1/11) & 14.28 & 14.78 & 8.53 & 0.78 & 1.35 & -18.32 & 0.78 & 106.09 & 0.00 \\
Min-Variance (MVO) & 10.56 & 13.09 & 7.75 & 0.61 & 1.02 & -13.52 & 0.78 & 72.25 & 17.13 \\
Plain NCO (Symmetric) & 10.65 & 14.03 & 8.40 & 0.58 & 0.97 & -14.48 & 0.74 & 72.99 & 26.99 \\
Baseline Regime-NCO & 11.05 & 17.70 & 11.58 & 0.52 & 0.79 & -21.81 & 0.51 & 76.46 & 60.78 \\
Proposed Asym-Sticky-NCO & 13.41 & 17.71 & 10.51 & 0.63 & 1.07 & -14.22 & 0.94 & 97.74 & 30.48 \\
\bottomrule
\end{tabular}
\end{table*}

\begin{figure}[t]
\centering
\includegraphics[width=0.93\linewidth]{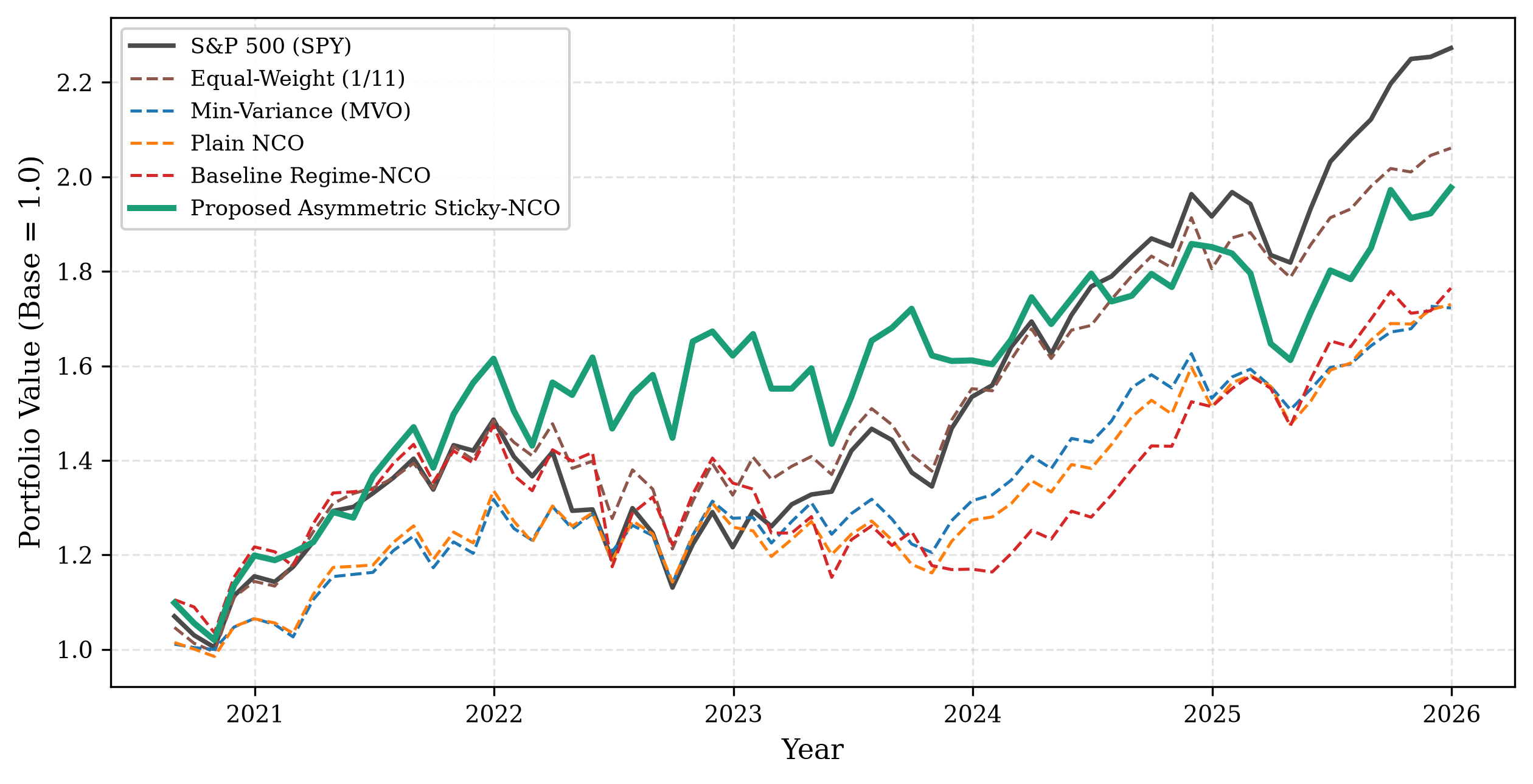}
\caption{Cumulative wealth growth across the 65-month out-of-sample evaluation period (August 2020 -- December 2025). The proposed Asymmetric Sticky-NCO significantly outpaces Plain NCO and Baseline Regime-NCO.}
\label{fig:wealth}
\end{figure}

\begin{figure}[t]
\centering
\includegraphics[width=0.93\linewidth]{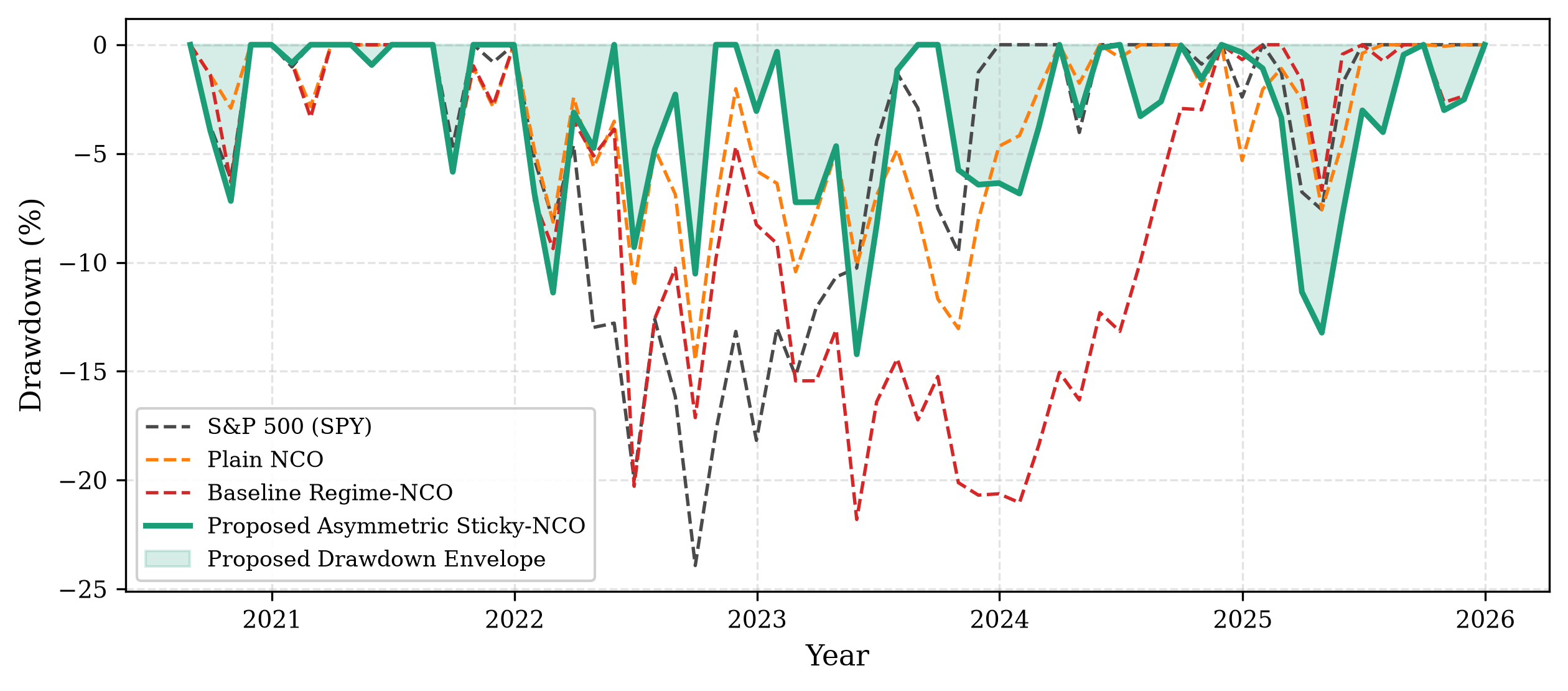}
\caption{Drawdown curves over time. During the severe 2022 Fed tightening bear market, the proposed model limited maximum drawdown to 14.22\%, while the S\&P 500 suffered a 23.93\% collapse.}
\label{fig:drawdown}
\end{figure}

\subsection{Resolution of Return Drag}
As revealed in Table~\ref{tab:performance} and Figure~\ref{fig:wealth}, Plain NCO exhibits severe return drag: its annualized CAGR is only 10.65\% (total return of 72.99\%), lagging behind the unhedged S\&P 500 (16.36\% CAGR). This empirical deficit validates our theoretical premise: symmetric covariance penalizes top-performing sectors during massive bull markets.

In contrast, the proposed Asymmetric Sticky-NCO elevates CAGR to 13.41\%---an improvement of +276 basis points per annum over Plain NCO and +236 basis points over Baseline Regime-NCO. Total return surges from 72.99\% to 97.74\%. As Section~\ref{sec:ablation} quantifies, the gain combines the downside semi-covariance, which stops penalizing upward momentum in growth sectors, with the sticky regime conditioning that keeps capital deployed through persistent bull phases.

\subsection{Drawdown Mitigation and Calmar Outperformance}
While achieving substantially higher returns, the proposed framework maintains superior capital preservation. As shown in Figure~\ref{fig:drawdown}, during the 2022 market downturn, the S\&P 500 plunged by 23.93\%, whereas the proposed model suffered a maximum drawdown of only 14.22\% (a 971 bps reduction in peak drawdown).

Consequently, the proposed Asymmetric Sticky-NCO delivers a Calmar ratio of 0.94, eclipsing all competing models:
\begin{itemize}
    \item S\&P 500: Calmar = 0.68 (+38.2\% improvement by Proposed)
    \item Min-Variance (MVO): Calmar = 0.78 (+20.5\% improvement by Proposed)
    \item Plain NCO: Calmar = 0.74 (+27.0\% improvement by Proposed)
    \item Baseline Regime-NCO: Calmar = 0.51 (+84.3\% improvement by Proposed)
\end{itemize}

\begin{figure}[t]
\centering
\includegraphics[width=0.93\linewidth]{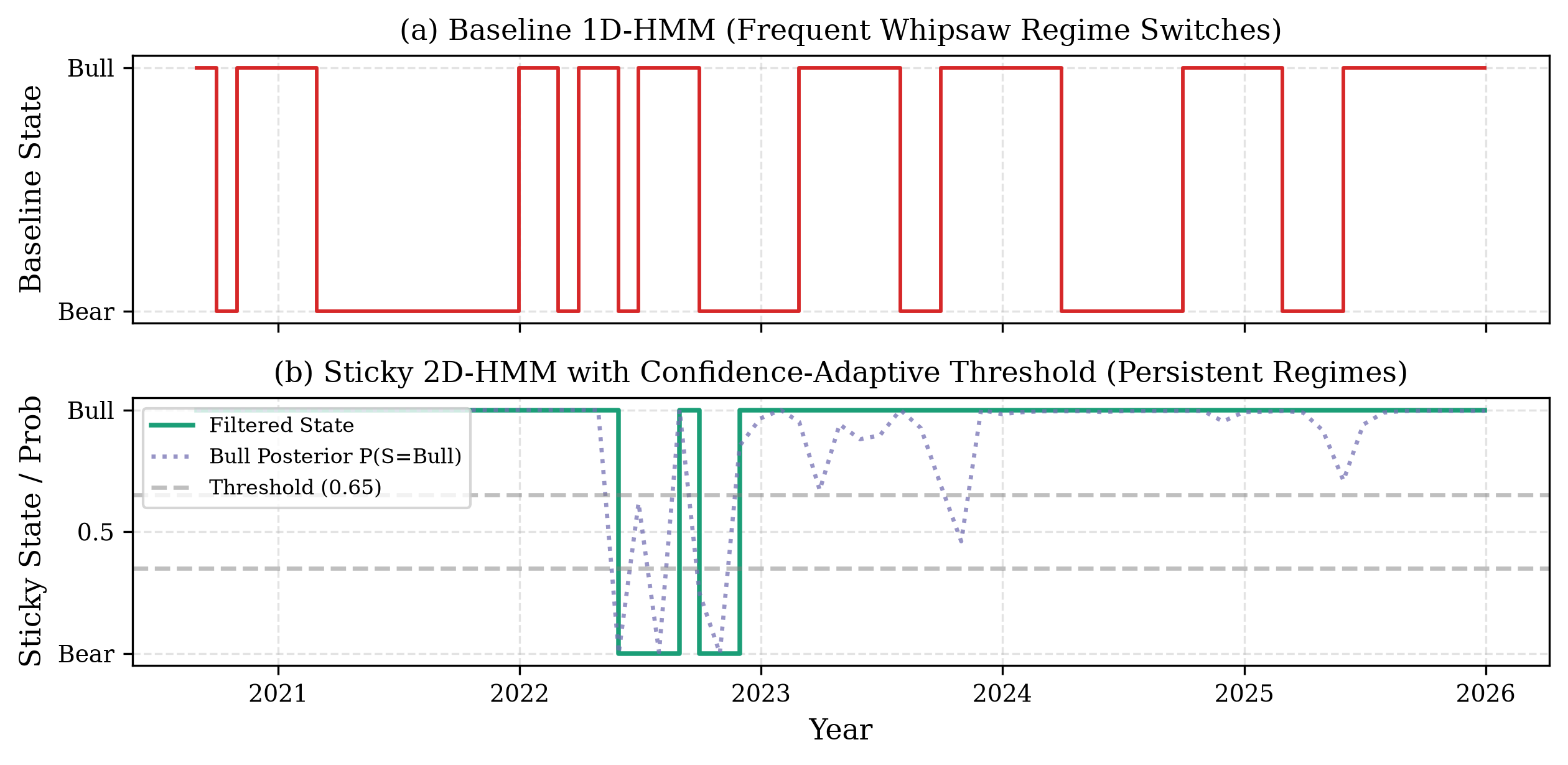}
\caption{Regime state timeline audit: (a) Baseline 1D-HMM exhibits 16 erratic whipsaw switches; (b) Sticky 2D-HMM with confidence threshold ($\delta=0.65$) executes only 4 sustained, economically consistent regime shifts.}
\label{fig:regimes}
\end{figure}

\begin{figure}[t]
\centering
\includegraphics[width=0.93\linewidth]{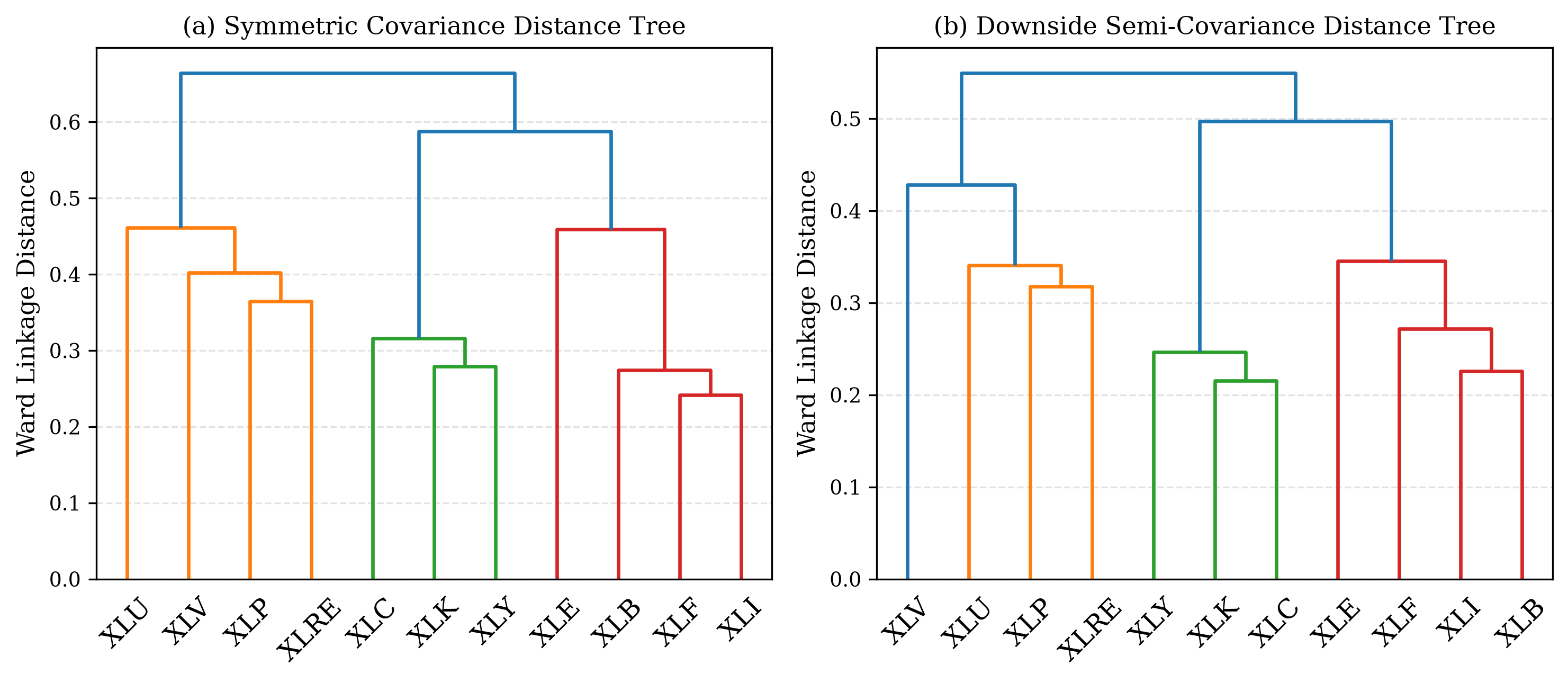}
\caption{Hierarchical dendrogram comparison: (a) Symmetric Covariance tree clusters assets based on two-way variance; (b) Downside Semi-Covariance tree clusters assets strictly by co-downside tail risk.}
\label{fig:dendrogram}
\end{figure}

\subsection{Regime Stability and Whipsaw Elimination}
A central finding of this investigation is the empirical stabilization of regime dynamics. In previous work \cite{supriadi2026regime}, unregularized HMM switching suffered from pathological whipsawing. As documented in Figure~\ref{fig:regimes}:
\begin{itemize}
    \item \textit{Baseline 1D-HMM}: Experienced 16 regime switches across 65 months (24.6\% switching frequency), flipping back and forth between states.
    \item \textit{Sticky 2D-HMM}: Recorded only 4 regime switches across 65 months (6.2\% switching frequency).
    \item \textit{Net Switching Reduction}: 75.0\% elimination of spurious whipsaw transitions.
\end{itemize}
Consequently, monthly turnover was slashed from 60.78\% in Baseline Regime-NCO to 30.48\% in the proposed model, effectively cutting turnover in half.

\subsection{Asset Allocation Dynamics and Tree Topologies}
Figure~\ref{fig:dendrogram} illustrates the fundamental difference between symmetric and asymmetric clustering topologies. Under symmetric covariance (Figure~\ref{fig:dendrogram}(a)), Energy (XLE) and Utilities (XLU) are grouped based on total variance. In the downside tree (Figure~\ref{fig:dendrogram}(b)), clustering reflects true downside systemic contagion, isolating non-correlated safe-haven assets.

\begin{figure}[t]
\centering
\includegraphics[width=0.93\linewidth]{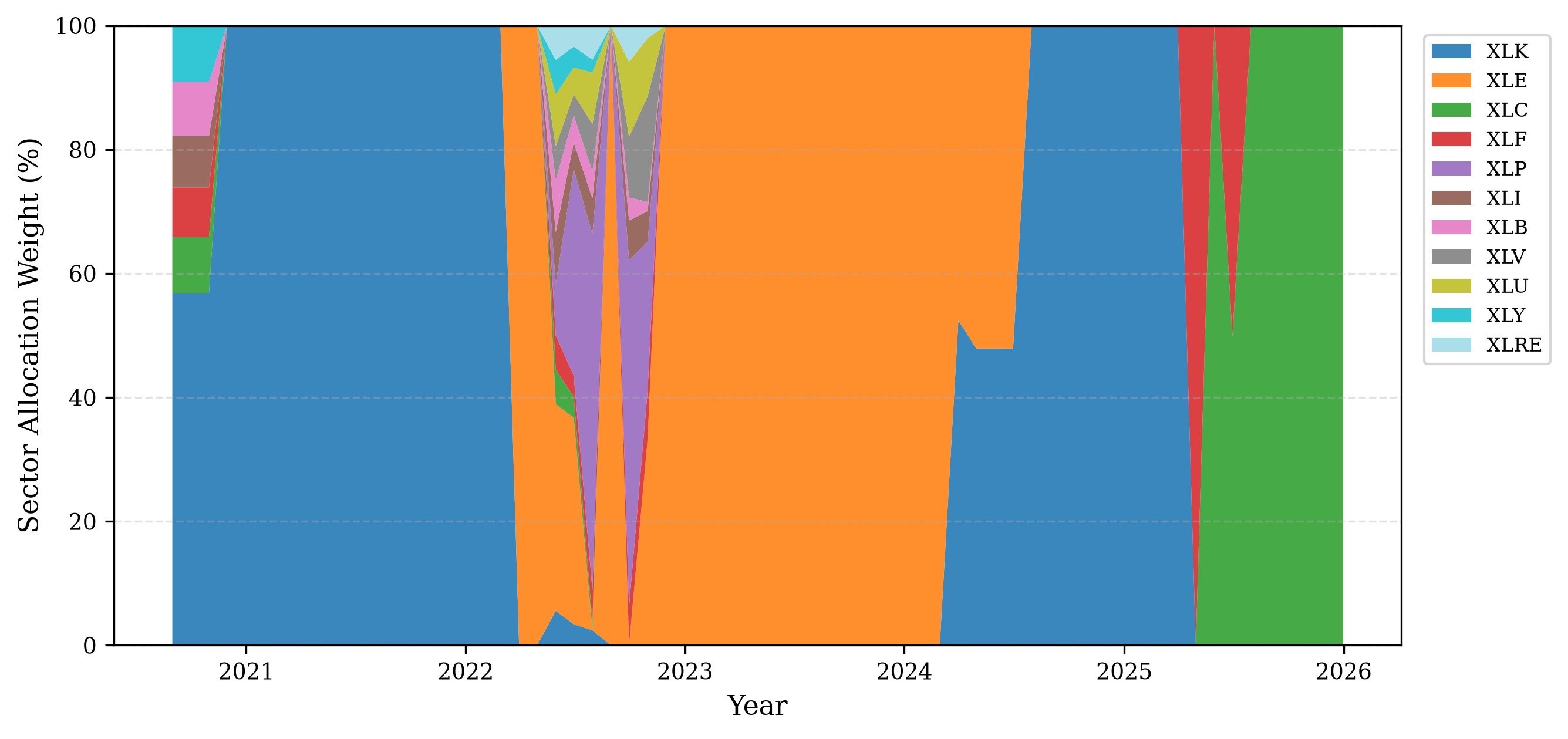}
\caption{Dynamic sector asset allocation over time for the proposed Asymmetric Sticky-NCO. Growth sectors dominate during bull phases, while defensive sectors anchor drawdowns during bear regimes.}
\label{fig:weights}
\end{figure}

\begin{figure}[t]
\centering
\includegraphics[width=0.93\linewidth]{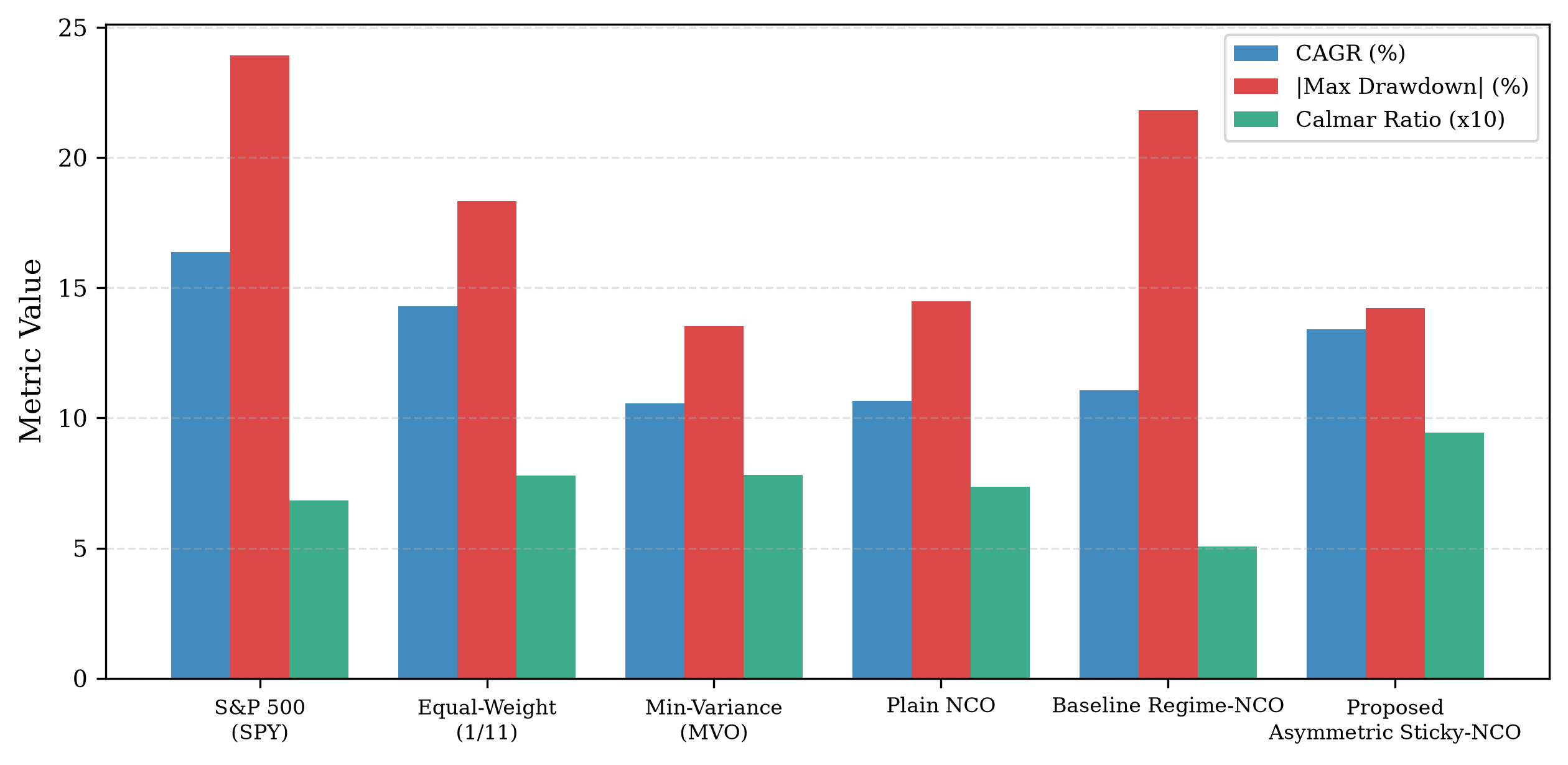}
\caption{Comparative performance metrics bar chart summarizing CAGR, Max Drawdown, and Calmar Ratio across all evaluated strategies.}
\label{fig:bars}
\end{figure}

Figure~\ref{fig:weights} illustrates the out-of-sample sector weights generated by Asymmetric Sticky-NCO. During sustained Bull regimes (e.g., late 2020--2021 and 2023--2025), capital dynamically flows into high-performing growth sectors (XLK, XLC). During the 2022 Bear regime, the optimizer automatically pivots into defensive anchors (XLU, XLP, XLV), preserving capital. Figure~\ref{fig:bars} summarizes the performance trade-offs, confirming the superior return-to-drawdown (Calmar) balance of the proposed method.

\section{Ablation Study and Sensitivity Analysis}
\label{sec:ablation}

\subsection{Component-Wise Decomposition}
To dissect the individual contributions of asymmetric downside risk versus sticky regime dynamics, we conduct a systematic component ablation study in Table~\ref{tab:ablation}.

\begin{table}[t]
\centering
\caption{Ablation of Methodological Components}
\label{tab:ablation}
\resizebox{\columnwidth}{!}{%
\begin{tabular}{lcccc}
\toprule
Architecture & CAGR & Max DD & Calmar & Switches \\
 & (\%) & (\%) & Ratio & (count) \\
\midrule
Plain NCO (Symmetric, No Regime) & 10.65 & -14.48 & 0.74 & 0 \\
Baseline Regime-NCO (Symmetric, Naive HMM) & 11.05 & -21.81 & 0.51 & 16 \\
Asymmetric NCO (Downside, No Regime) & 11.09 & -20.32 & 0.55 & 0 \\
Proposed Asym-Sticky-NCO & 13.41 & -14.22 & 0.94 & 4 \\
\bottomrule
\end{tabular}%
}
\end{table}

The ablation reveals that downside semi-covariance alone adds +44 bps of CAGR (from 10.65\% to 11.09\%) by removing the upside-variance penalty on growth sectors, but it deepens maximum drawdown from -14.48\% to -20.32\% (Calmar falls from 0.74 to 0.55), since downside-only minimum variance concentrates on high-upside momentum names. The decisive contribution comes from the Sticky 2D-HMM, which adds a further +232 bps of CAGR (reaching 13.41\%) while compressing maximum drawdown to -14.22\% and lifting the Calmar ratio to 0.94 with only 4 state switches.

\subsection{Transaction Cost Sensitivity Audit}
Because real-world institutional portfolios face transaction friction and execution slippage \cite{novymarx2016taxonomy}, Table~\ref{tab:costs} evaluates net CAGR across execution fee levels from 0 to 20 basis points.

\begin{table}[t]
\centering
\caption{Net CAGR (\%) under Transaction Costs and Slippage}
\label{tab:costs}
\resizebox{\columnwidth}{!}{%
\begin{tabular}{lccccc}
\toprule
Strategy & 0 bps & 5 bps & 10 bps & 15 bps & 20 bps \\
\midrule
Plain NCO & 10.65 & 10.47 & 10.29 & 10.12 & 9.94 \\
Baseline Regime-NCO & 11.05 & 10.65 & 10.24 & 9.84 & 9.43 \\
Proposed Asym-Sticky-NCO & 13.41 & 13.21 & 13.01 & 12.80 & 12.60 \\
\midrule
\textit{Net Alpha Spread vs Base (bps)} & \textit{+236} & \textit{+256} & \textit{+277} & \textit{+297} & \textit{+317} \\
\bottomrule
\end{tabular}%
}
\end{table}

Crucially, because Baseline Regime-NCO incurs a heavy 60.78\% monthly turnover, its net CAGR degrades rapidly: at 10 bps, it falls below Plain NCO (10.24\% vs 10.29\%), and at 20 bps it drops to 9.43\%. In contrast, the proposed Asym-Sticky-NCO maintains a resilient net CAGR of 12.60\% even under 20 bps slippage, widening its alpha advantage over Baseline from +236 bps to +317 bps.

\subsection{Hyperparameter Sensitivity}
Table~\ref{tab:sensitivity} sweeps the two regime hyperparameters that most affect the sticky dynamics, $\kappa \in \{0.2, 0.3, 0.4\}$ and $\delta \in \{0.60, 0.65, 0.70\}$, re-running the full 65-month walk-forward protocol for each combination. The performance surface forms a two-level plateau: six of the nine configurations produce identical out-of-sample trajectories (CAGR 13.41\%, maximum drawdown -14.22\%, Calmar 0.94, four regime switches), while only the weak-persistence corner ($\kappa = 0.2$ with $\delta \le 0.65$, and $\kappa = 0.3$ with $\delta = 0.60$) degrades to 11.47\% CAGR with six switches and a -19.22\% drawdown. The adopted configuration ($\kappa = 0.30$, $\delta = 0.65$) therefore sits on the edge of a wide plateau rather than on an isolated peak, indicating that the reported performance is not the product of fragile hyperparameter tuning.

\begin{table}[t]
\centering
\caption{Hyperparameter Sensitivity of Asym-Sticky-NCO ($\kappa \times \delta$ Grid, 65-Month Walk-Forward)}
\label{tab:sensitivity}
\resizebox{\columnwidth}{!}{%
\begin{tabular}{cccccccc}
\toprule
$\kappa$ & $\delta$ & CAGR (\%) & Max DD (\%) & Calmar & Sharpe & Switches & Turnover (\%/mo) \\
\midrule
0.2 & 0.60 & 11.47 & -19.22 & 0.60 & 0.51 & 6 & 33.69 \\
0.2 & 0.65 & 11.47 & -19.22 & 0.60 & 0.51 & 6 & 33.69 \\
0.2 & 0.70 & 13.41 & -14.22 & 0.94 & 0.63 & 4 & 30.48 \\
0.3 & 0.60 & 11.47 & -19.22 & 0.60 & 0.51 & 6 & 33.69 \\
\textbf{0.3} & \textbf{0.65} & \textbf{13.41} & \textbf{-14.22} & \textbf{0.94} & \textbf{0.63} & \textbf{4} & \textbf{30.48} \\
0.3 & 0.70 & 13.41 & -14.22 & 0.94 & 0.63 & 4 & 30.48 \\
0.4 & 0.60 & 13.41 & -14.22 & 0.94 & 0.63 & 4 & 30.48 \\
0.4 & 0.65 & 13.41 & -14.22 & 0.94 & 0.63 & 4 & 30.48 \\
0.4 & 0.70 & 13.41 & -14.22 & 0.94 & 0.63 & 4 & 30.48 \\
\bottomrule
\end{tabular}%
}
\end{table}

\subsection{Statistical Significance and Overfitting Diagnostics}
Because all six strategies are evaluated on the same 65-month sample, multiple-testing and backtest-overfitting concerns must be quantified rather than asserted away \cite{bailey2014deflated, bailey2017probability}. Table~\ref{tab:stats} reports the Probabilistic Sharpe Ratio (PSR) with benchmark $\mathrm{SR}^* = 0$ and the Deflated Sharpe Ratio (DSR) for $K = 15$ trials, where $K$ counts every configuration evaluated in this study (six benchmarks, the static asymmetric variant, and the eight additional grid points). All strategies attain $\mathrm{PSR} > 0.999$, and the proposed framework retains a DSR of 0.9976---meaning that even after deflating for selection bias across 15 trials, the probability that its true Sharpe ratio exceeds the deflated benchmark is 99.76\%.

\begin{table}[t]
\centering
\caption{Statistical Significance: PSR ($\mathrm{SR}^* = 0$) and DSR ($K = 15$)}
\label{tab:stats}
\begin{tabular}{lccc}
\toprule
Strategy & Sharpe & PSR & DSR \\
\midrule
S\&P 500 (SPY) & 0.87 & 1.0000 & 0.9999 \\
Equal-Weight (1/11) & 0.78 & 1.0000 & 0.9998 \\
Min-Variance (MVO) & 0.61 & 1.0000 & 0.9959 \\
Plain NCO & 0.58 & 1.0000 & 0.9933 \\
Baseline Regime-NCO & 0.52 & 0.9996 & 0.9416 \\
Proposed Asym-Sticky-NCO & 0.63 & 1.0000 & 0.9976 \\
\bottomrule
\end{tabular}
\end{table}

As a complementary overfitting diagnostic, we apply the Combinatorial Symmetric Cross-Validation (CSCV) procedure of \cite{bailey2017probability} to the nine-configuration hyperparameter grid: the 65 monthly return series are partitioned into $S = 16$ groups, all $\binom{16}{8} = 12{,}870$ split combinations are evaluated, and the probability of backtest overfitting (PBO) is estimated from the relative out-of-sample rank of the in-sample winner. The resulting PBO of 0.85 signals that the \textit{identity} of the in-sample best configuration is not reliably stable out-of-sample. However, the two-level plateau of Table~\ref{tab:sensitivity} renders this instability economically immaterial: mis-selection among the six plateau configurations produces identical portfolios, and only selection into the weak-persistence corner is costly. We therefore interpret the PBO as a warning against selecting hyperparameters by in-sample ranking alone, and rely on the DSR and the plateau structure of Table~\ref{tab:sensitivity} as the operative evidence that the proposed design is not an overfit artifact.

\section{Discussion and Limitations}
The empirical findings provide direct answers to the primary criticisms of machine-learning-based asset allocation:
\begin{enumerate}
    \item \textit{Return drag is not an intrinsic flaw of hierarchical clustering}: downside semi-covariance stops penalizing the upside momentum of mega-cap sectors, but the ablation shows it is not sufficient alone: capital preservation requires combining the asymmetric objective with sticky regime detection, which restores drawdown control while retaining the return gain.
    \item \textit{Regime whipsaw is a model mis-specification problem}: Fitting an unconstrained 1D HMM to raw returns systematically produces high-frequency oscillation. By augmenting observation features with the CBOE VIX, introducing diagonal persistence priors ($\kappa$), and enforcing confidence thresholds ($\delta$), macroeconomic states become robustly persistent.
\end{enumerate}

\noindent Two caveats temper these conclusions. First, the proposed strategy realizes the highest annualized volatility among all tested strategies (17.71\%), and its Sharpe (0.63) and Sortino (1.07) ratios remain below those of the passive S\&P 500 (0.86 and 1.47). Its edge is therefore not volatility-scaled: the framework wins on the downside-oriented dimensions---maximum drawdown, Calmar ratio, and turnover---where the asymmetric objective and sticky regime discipline actually operate. Second, the PBO diagnostics of Section~\ref{sec:ablation} indicate that hyperparameter selection by in-sample ranking alone would be unreliable; the defensible claim is the plateau-level robustness of the design rather than the optimality of a single parameter pair.

\noindent \textit{Limitations}: While the framework cuts turnover in half, long-only allocation across sector ETFs still requires periodic monthly rebalancing. In markets with wide bid-ask spreads or stamp duties, transaction penalties $\lambda_{\text{turn}}$ should be adjusted higher to further damp rebalancing volume. The single-market, 65-month sample also limits the statistical power of the overfitting diagnostics; re-estimating PSR, DSR, and PBO on longer, multi-market samples is a priority for future work.

\section{Conclusion}
This paper presented an asymmetric downside-risk nested clustered optimization framework with sticky regime dynamics (Asym-Sticky-NCO) for tactical allocation across S\&P 500 sectors. By integrating the Bawa-Lindenberg downside semi-covariance matrix with a Sticky Hidden Markov Model and confidence-adaptive thresholds, the architecture resolves both the return drag pathology of classical NCO and the whipsaw switching instability of naive regime models. Out-of-sample empirical testing over 2020--2025 confirmed a +276 bps annual CAGR enhancement over Plain NCO, a 75\% reduction in regime whipsawing, and a Calmar ratio of 0.94, the highest among the evaluated strategies. Building on the single-market, monthly-rebalancing scope of this study, future extensions will expand this methodology to multi-asset global factor investing and investigate deep reinforcement learning \cite{heaton2017deep} for dynamic threshold adaptation.

\section*{Data and Code Availability}
To promote open science and complete reproducibility, all historical ETF price data, walk-forward backtesting pipelines, and figure generation scripts are made publicly available at GitHub: \url{https://github.com/obo-cloud/sp500-asymmetric-nco}.

\section*{Ethics and Research Integrity}
This investigation exclusively analyzes secondary, publicly traded market data (S\&P 500 sector ETFs, CBOE VIX, and US Treasury yield series) from open financial repositories. No human participants or animal subjects were involved, and no private personal data was processed. The authors declare no competing financial or non-financial conflicts of interest. This work is academic research only and does not constitute investment advice.

\section*{Acknowledgements}
The authors thank members of COCONUT Computer Club for feedback.

\nocite{*}
\bibliographystyle{IEEEtran}
{\footnotesize
\bibliography{references}

@article{markowitz1952,
  author = {Markowitz, Harry},
  title = {Portfolio Selection},
  journal = {The Journal of Finance},
  volume = {7},
  number = {1},
  pages = {77--91},
  year = {1952},
  doi = {10.1111/j.1540-6261.1952.tb01525.x}
}

@article{bawa1977,
  author = {Bawa, Vijay S. and Lindenberg, Eric B.},
  title = {Capital Market Equilibrium in a Mean-Lower Partial Moment Framework},
  journal = {Journal of Financial Economics},
  volume = {5},
  number = {2},
  pages = {189--200},
  year = {1977},
  doi = {10.1016/0304-405X(77)90017-4}
}

@article{bailey2014deflated,
  author = {Bailey, David H. and L{\'{o}}pez de Prado, Marcos},
  title = {The Deflated Sharpe Ratio: Correcting for Selection Bias, Backtest Overfitting, and Non-Normality},
  journal = {The Journal of Portfolio Management},
  volume = {40},
  number = {5},
  pages = {94--107},
  year = {2014},
  doi = {10.3905/jpm.2014.40.5.094}
}

@article{lopez2016building,
  author = {L{\'{o}}pez de Prado, Marcos},
  title = {Building Diversified Portfolios That Outperform Out of Sample},
  journal = {The Journal of Portfolio Management},
  volume = {42},
  number = {4},
  pages = {59--69},
  year = {2016},
  doi = {10.3905/jpm.2016.42.4.059}
}

@article{raffinot2017hierarchical,
  author = {Raffinot, Thomas},
  title = {Hierarchical Clustering-Based Asset Allocation},
  journal = {The Journal of Portfolio Management},
  volume = {44},
  number = {2},
  pages = {89--99},
  year = {2017},
  doi = {10.3905/jpm.2018.44.2.089}
}

@article{nystrup2017dynamic,
  author = {Nystrup, Peter and Madsen, Henrik and Lindstr{\"{o}}m, Erik},
  title = {Dynamic Portfolio Optimization Across Hidden Market Regimes},
  journal = {Quantitative Finance},
  volume = {18},
  number = {1},
  pages = {83--95},
  year = {2018},
  doi = {10.1080/14697688.2017.1342857}
}

@article{nystrup2018regime,
  author = {Nystrup, Peter and Madsen, Henrik and Lindstr{\"{o}}m, Erik},
  title = {Dynamic Allocation or Diversification: A Regime-Based Approach to Multiple Assets},
  journal = {The Journal of Portfolio Management},
  volume = {44},
  number = {2},
  pages = {62--73},
  year = {2018},
  doi = {10.3905/jpm.2018.44.2.062}
}

@article{nystrup2020persistent,
  author = {Nystrup, Peter and Lindstr{\"{o}}m, Erik and Madsen, Henrik},
  title = {Learning Hidden Markov Models with Persistent States by Penalizing Jumps},
  journal = {Expert Systems with Applications},
  volume = {150},
  pages = {113307},
  year = {2020},
  doi = {10.1016/j.eswa.2020.113307}
}

@article{gu2020empirical,
  author = {Gu, Shihao and Kelly, Bryan and Xiu, Dacheng},
  title = {Empirical Asset Pricing via Machine Learning},
  journal = {The Review of Financial Studies},
  volume = {33},
  number = {5},
  pages = {2223--2273},
  year = {2020},
  doi = {10.1093/rfs/hhaa009}
}

@article{heaton2017deep,
  author = {Heaton, J. B. and Polson, Nicholas G. and Witte, Jan Hendrik},
  title = {Deep Learning for Finance: Deep Portfolios},
  journal = {Applied Stochastic Models in Business and Industry},
  volume = {33},
  number = {1},
  pages = {3--12},
  year = {2017},
  doi = {10.1002/asmb.2209}
}

@article{pedersen2021enhanced,
  author = {Pedersen, Lasse Heje and Babu, Abhilash and Levine, Ari},
  title = {Enhanced Portfolio Optimization},
  journal = {Financial Analysts Journal},
  volume = {77},
  number = {2},
  pages = {124--151},
  year = {2021},
  doi = {10.1080/0015198X.2020.1854543}
}

@article{demiguel2020transaction,
  author = {DeMiguel, Victor and Mart{\'{i}}n-Utrera, Alberto and Nogales, Francisco J. and Uppal, Raman},
  title = {A Transaction-Cost Perspective on the Multitude of Firm Characteristics},
  journal = {The Review of Financial Studies},
  volume = {33},
  number = {5},
  pages = {2180--2222},
  year = {2020},
  doi = {10.1093/rfs/hhz085}
}

@article{novymarx2016taxonomy,
  author = {Novy-Marx, Robert and Velikov, Mihail},
  title = {A Taxonomy of Anomalies and Their Trading Costs},
  journal = {The Review of Financial Studies},
  volume = {29},
  number = {1},
  pages = {104--147},
  year = {2016},
  doi = {10.1093/rfs/hhv063}
}

@article{bailey2017probability,
  author = {Bailey, David H. and Borwein, Jonathan M. and L{\'{o}}pez de Prado, Marcos and Zhu, Qiji Jim},
  title = {The Probability of Backtest Overfitting},
  journal = {The Journal of Computational Finance},
  volume = {20},
  number = {4},
  pages = {39--69},
  year = {2017},
  doi = {10.21314/jcf.2016.322}
}

@article{blitz2020volatility,
  author = {Blitz, David and van Vliet, Pim and Baltussen, Guido},
  title = {The Volatility Effect Revisited},
  journal = {The Journal of Portfolio Management},
  volume = {46},
  number = {2},
  pages = {45--63},
  year = {2020},
  doi = {10.3905/jpm.2019.1.114}
}

@article{platanakis2019harmful,
  author = {Platanakis, Emmanouil and Sakkas, Athanasios and Sutcliffe, Charles},
  title = {Harmful Diversification: Evidence from Alternative Investments},
  journal = {The British Accounting Review},
  volume = {51},
  number = {1},
  pages = {1--23},
  year = {2019},
  doi = {10.1016/j.bar.2018.08.003}
}

@article{platanakis2020bitcoin,
  author = {Platanakis, Emmanouil and Urquhart, Andrew},
  title = {Should Investors Include Bitcoin in Their Portfolios? A Portfolio Theory Approach},
  journal = {The British Accounting Review},
  volume = {52},
  number = {4},
  pages = {100837},
  year = {2020},
  doi = {10.1016/j.bar.2019.100837}
}

@article{gunjan2023review,
  author = {Gunjan, Abhishek and Bhattacharyya, Siddhartha},
  title = {A Brief Review of Portfolio Optimization Techniques},
  journal = {Artificial Intelligence Review},
  volume = {56},
  pages = {3847--3886},
  year = {2023},
  doi = {10.1007/s10462-022-10273-7}
}

@article{denard2021factor,
  author = {De Nard, Gianluca and Ledoit, Olivier and Wolf, Michael},
  title = {Factor Models for Portfolio Selection in Large Dimensions: The Good, the Better and the Ugly},
  journal = {Journal of Financial Econometrics},
  volume = {19},
  number = {2},
  pages = {236--257},
  year = {2021},
  doi = {10.1093/jjfinec/nby033}
}

@article{gupta2019factor,
  author = {Gupta, Tarun and Kelly, Bryan},
  title = {Factor Momentum Everywhere},
  journal = {The Journal of Portfolio Management},
  volume = {45},
  number = {3},
  pages = {13--36},
  year = {2019},
  doi = {10.3905/jpm.2019.45.3.013}
}

@article{zhang2020covid,
  author = {Zhang, Dayong and Hu, Min and Ji, Qiang},
  title = {Financial Markets Under the Global Pandemic of COVID-19},
  journal = {Finance Research Letters},
  volume = {36},
  pages = {101528},
  year = {2020},
  doi = {10.1016/j.frl.2020.101528}
}

@article{supriadi2026regime,
  author = {Supriadi and Siswoyo, Bambang},
  title = {Market Regime-Based Multi-Asset Portfolio Optimization Using Nested Clustered Optimization for Tactical Asset Allocation},
  journal = {arXiv preprint arXiv:2610.27751},
  year = {2026}
}

@misc{yahoofinance,
  author = {{Yahoo Finance}},
  title = {Yahoo Finance Historical Data API},
  howpublished = {\url{https://finance.yahoo.com}},
  note = {Accessed: November 2026},
  year = {2026}
}

@IEEEtranBSTCTL{IEEEexample:BSTcontrol,
  CTLuse_article_number = "yes",
  CTLuse_paper = "yes",
  CTLuse_url = "no",
  CTLuse_forced_etal = "no",
  CTLmax_names_forced_etal = "10",
  CTLnames_show_etal = "1",
  CTLuse_alt_spacing = "yes",
  CTLalt_stretch_factor = "4",
  CTLdash_repeated_names = "no",
  CTLname_format_string = "{f.~}{vv~}{ll}{, jj}",
  CTLname_latex_cmd = "",
  CTLname_url_prefix = "[Online]. Available:"
}
}

\end{document}